\documentclass[prl,twocolumn,showpacs]{revtex4}

\usepackage{graphicx} 
\usepackage{bm}       
\usepackage{dcolumn}  
\usepackage{amsmath}

\newcommand{\bk}{\bm{k}}

\begin{document}

\title{
Quantum Hall effects of graphene with multi orbitals:\\
Topological numbers, Boltzmann conductance and  Semi-classical quantization
}

\author{Masao Arai}
\email{arai.masao@nims.go.jp}
\affiliation{
  Computational Materials Science Center, National Institute for Materials Science, 
  Tsukuba, Ibaraki 305-0044, Japan }
\author{Yasuhiro Hatsugai}
\email{hatsugai@sakura.cc.tsukuba.ac.jp}
\affiliation{
  Institute of Physics, University of Tsukuba, Tsukuba, Ibaraki, 305-8571, Japan }

\date{\today}

\begin{abstract} 

Hall conductance $\sigma_{xy}$ 
as the Chern numbers of the Berry connection in the magnetic Brillouin zone
is calculated for a realistic multi band tight-band 
model of graphene with non-orthogonal basis. 
It is confirmed that the envelope of $\sigma_{xy}$ coincides with  
a semi-classical result when magnetic field is sufficiently small.
 The Hall resistivity $\rho_{xy}$ from the 
 weak-field Boltzmann theory also explains
the overall behaviour of the  $\sigma_{xy}$ if the Fermi surface is composed of
a single energy band. The plateaux of $\sigma_{xy}$ are explained 
from semi-classical quantization 
and necessary modification  is proposed for the Dirac fermion regimes.

\end{abstract}

\pacs{73.43.-f, 73.21.-b, 03.65.Sq}

\maketitle

The quantum Hall effect (QHE) has been 
 one of the important subjects in condensed matter physics 
for several decades. Wide range of experimental and theoretical studies have been 
devoted to this subject.
In the early stage of these efforts, the topological aspects of quantization have been firmly established\cite{thouless1982,streda1982,kohmoto1985,niu1985,aoki1986,halperin1982,hatsugai1993}. 
The topological formulation of 
quantized Hall conductance\cite{fukui2005} has been utilized numerically
to explore exotic nature of electrons on two-dimensional lattice models.

Recently, electrons on a honeycomb lattice has attracted theoretical attention 
because energy bands on this lattice are described by massless Dirac dispersion
near the center of bands. The dispersion results in anomalous 
QHE $\sigma_{xy} = (2N+1)e^2/h$ ($N$: integer)
per spin\cite{zheng2002,gusynin2005}.
Such anomalous QHE was observed experimentally in graphene, which is monolayer graphite\cite{novoselov2005,zhang2005}. The topological aspect of QHE in graphene has been 
studied by using a single-orbital tight-binding model on a honeycomb lattice\cite{sheng2006,hatsugai2006}.
It has been shown that anomalous QHE persists up to the van Hove singularities\cite{hatsugai2006}.
At the singularities, $\sigma_{xy}$ jumps discontinuously and normal QHE $\sigma_{xy} = N e^2/h$ 
is recovered around band edge regions. These results suggest that bands structure affects 
QHE both qualitatively and quantitatively. Thus, it is an interesting question to investigate 
the relation between band structure and $\sigma_{xy}$.

In this letter, we study relations between band structure and quantized $\sigma_{xy}$ 
by using a realistic tight-binding model of graphene as an example.
Up to now, numerical calculations of  quantized $\sigma_{xy}$ have been rather limited to simple 
tight-binding models and realistic band structure was beyond scopes of these studies.
Another purpose of this study is to demonstrate that the topological formulation of $\sigma_{xy}$\cite{thouless1982,fukui2005} can be 
applicable to realistic band structures.

We use a tight-binding model for graphene with $s$ and $p$ orbitals on carbon atoms\cite{saito1992}.
The spin degeneracy is ignored for simplicity.
Each orbital $| i \xi \rangle$ is labeled by a site index $i$ and  an orbital index $\xi$. 
Transfer $t_{i\xi, j\xi'}$ 
and overlap integrals $s_{i\xi, j\xi'}$ between nearest-neighbor atoms are considered by a 
Slater-Koster approximation.The onsite
energy $\varepsilon_{i\xi}$, $t_{i\xi, j\xi'}$ and $s_{i\xi, j\xi'}$ 
are obtained from the  parameters in Refs. \onlinecite{saito1992} and \onlinecite{min2006}. 
With eight orbitals in a unit cell,
energy bands dispersion is calculated as shown in Fig. 1. In actual graphene,  
electrons are occupied up to $\varepsilon = 0$. Around this energy, band dispersions are
well approximated by the massless Dirac cones, which is the origin of several interesting physical
properties including anomalous quantization of Hall conductance.
The massless Dirac dispersions are also realized around $\varepsilon \sim -14\text{eV}$. 
In addition,
several van Hove singularities are clearly seen in the total density of states (DOS).

The uniform magnetic field is introduced as 
\begin{align}
  \left\langle i \xi \right| \hat{H} \left| j \xi' \right\rangle &= e^{i\theta_{ij}} t_{i\xi, j\xi'}, \ \
  \left\langle i \xi | j \xi' \right\rangle = e^{i\theta_{ij}} s_{i\xi, j\xi'}
\end{align}
so that they satisfy

$  \sum_{\text{closed loop}} \theta_{ij} = 2\pi\text{(flux quanta in the loop)}$.

Under the uniform magnetic field, original translational symmetry is broken. The system
is then characterized by the magnetic flux $\phi = B\Omega/\varphi_0$ where $\Omega$ is the
area of a unit cell and $\varphi_0 = hc/e$ is the quantized magnetic flux.
When $\phi$ is a rational number
$p/q$, we can define magnetic unit cell whose area is $q$ times
larger than that of the original cell.  Then, extended Bloch
theorem holds for the magnetic unit cell and eigenstates are
labeled by a wave-number $\bk$. Thus, an energy band
splits into $q$ sub-bands by the uniform field. 
If the chemical potential is located within an energy gap,
quantized Hall conductance $\sigma_{xy}$ is given as
\begin{equation}
  \sigma_{xy} = -\frac{e^2}{h} c_F(\mu),
\end{equation}
where $c_F$ is a topological integer called Chern number defined 
for the filled bands.
It is calculated 
by discretizing the Brillouin zone into mesh
$\{\bk _\ell\}$  as\cite{thouless1982,kohmoto1985,fukui2005}
\begin{eqnarray*}
c_F(\mu) &=&
\frac {1}{2\pi } \sum _\ell F(\bk_\ell) 
\\
F(\bk) &=&\text{Arg}\, u_x(\bk)u_y(\bk+\Delta \bk_x)
[u_x(\bk+\Delta\bk_y)u_y(\bk)] ^*
\\
u_\mu(\bk)  &=&\det [\Psi(\bk) ] ^\dagger \Psi(\bk+\Delta\bk_\mu)
\\
\Psi(\bk)  &=&(|\psi_1(\bk) \rangle ,\cdots, |\psi_M(\bk)\rangle )
\end{eqnarray*}
where 
 $\Delta \bk_\nu$ 
is a discretized momentum along $\nu$ direction,
 $| \psi_m(\bk) \rangle $ is a Bloch state of the band index $m$ 
and $m=M$ is the highest occupied energy band below 
the chemical potential $\mu$.
(See Ref.\cite{fukui2005} for the detail).

This formulation was successfully applied to quantum Hall effect on a
disorder system\cite{song2007} and graphene\cite{hatsugai2006}.
When the  field $\phi=p/q$ is sufficiently small, $p$ sub-bands are
grouped together in general. These grouped bands correspond to a Landau band 
of nearly free electrons.
Hereafter, we treat the  field $\phi=1/{q}$, 
which enables us to consider the weak-field limit easily.

\begin{figure}
  \begin{center}

  \includegraphics[width=8cm]{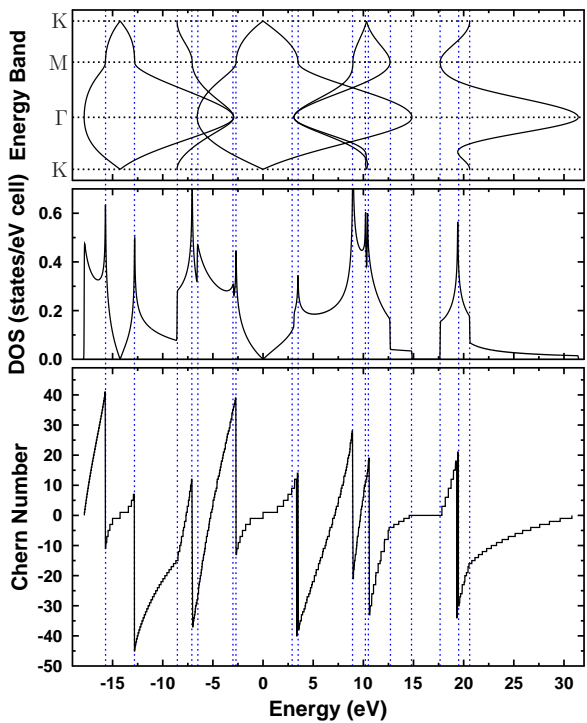}
  \end{center}
 \caption{(Color online) 
 The lower panel shows the calculated Chern number
   ($\sigma_{xy}$) for $\phi = 1/53$ as a function of the chemical
   potential. The upper and middle panels present
   energy bands and total density of states  when
    magnetic field is absent. Vertical dashed lines indicate positions of van Hove singularities.
   }
  \label{fig:sxy}
\end{figure}

In Fig.~\ref{fig:sxy}, we show the calculated Chern number $c_F$ with $\phi =
1/53$ as a function of the chemical potential.
The $c_F$ of adjacent gaps are connected by a straight
line as an eye guide.
In some energy regions, several Landau bands are grouped together
with very small energy gaps. In such case, we found that Chern numbers 
corresponding  to these small gaps are sensitive to the choice of $\phi$.
Thus, they are not physically relevant and we omit them from Fig.~\ref{fig:sxy}.
The  plateaux are most clearly seen at Dirac fermion regions because of 
large energy gaps between Landau bands. 
The $c_F$ increases 1, 2, or 3 between adjacent (well-developed) energy gaps.  
If these steps are
smeared, coarse grained $\sigma_{xy}$ behaves as a continuous function except
for several energies where the $c_F$ changes with large numbers.

We found that the large discontinuities in $c_F$ can be attributed to
topological changes of Fermi surface when the magnetic field is
absent. For example, the discontinuity at $\varepsilon \approx -15.7
\text{eV}$ is caused by a transformation of Fermi surface from an
electron-pocket  around $\Gamma$ point to two hole-pockets around K points
(see Fig.~\ref{fig:sxy2} (a) and (b)).

Such topological changes are also recognized as a van Hove singularity in
DOS. Other discontinuities
also correspond to similar topological changes of Fermi
surface.

\begin{figure}
  \begin{center}

  \includegraphics[width=8cm]{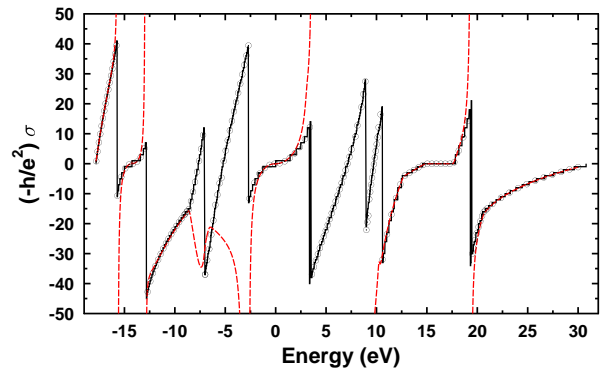}
  \end{center}
 \caption{(Color online)
 Comparison of the Hall conductance obtained from Chern number with semi-classical
 formulations. The solid line shows the $\sigma_{xy}$ obtained from Chern number and open circles
 indicate that from semi-classical theory with clean limit. The dashed line is $-\rho_{xy}^{-1}$ obtained
 from weak-field limit of Boltzmann theory.}
  \label{fig:sc}
\end{figure}

We compare global behaviour of $\sigma_{xy}$ with a semi-classical theory.
When Fermi surface consists of closed curves, the Hall
conductance $\sigma_{xy}$ for clean-limit can be calculated as\cite{lifshitz1956,lifshitz1957}

\begin{equation} 
  \sigma^\text{sc}_{xy} (\mu) = -\frac{ec}{(2\pi)^2 B} \sum_i S_i(\mu)
\end{equation}

within semi-classical approximation.
Here, $S_i(\mu)$ is the directed area enclosed by $i$-th segment of
Fermi surface.  It is defined as a
positive (negative) value when the area enclosed by the segment is
filled (empty). Above equation can be casted to the form:

\begin{equation} 
  \sigma^\text{sc}_{xy} (\mu) =   -\frac{e^2}{h} c_\text{sc}(\mu); \;
  c_\text{sc} (\mu) = \frac{\sum_i S_i(\mu)}{\phi\Omega_{\text{BZ}}}, 
\end{equation}

where $\Omega_{\text{BZ}} = (2\pi)^2/\Omega$ is the area of Brillouin zone. We found
that $c_\text{sc}$ completely describes the envelope behavior of
topological number $c_F$ for weak field limit as shown in Fig.~\ref{fig:sc}.
While this result
may be naively expected as we compare same physical quantity $\sigma_{xy}$,
it is highly non-trivial from the fully quantum mechanical formulation of Hall conductance. 
We have succeeded in demonstrating numerically that quantized Hall conductance
agrees with semi-classical limit when the applied field is sufficiently small.
Conversely, this agreement indicates that
the formulation based on Chern number is quantitatively applicable
to calculate $\sigma_{xy}$ in clean-limit for realistic energy bands. 
Thus, this formula may be useful to explore quantum Hall effect for materials with
exotic energy band structures.

The topological changes of Fermi surface occur as transitions
from electron pockets to hole pockets.
At these points, $\sum_i S_i$ discontinuously decreases with $\Omega_{BZ}$. 
Therefore,  discontinuity of Chern number is $\phi^{-1} = q$ and Hall 
conductance jumps by $  (-e^2/h)\phi^{-1}$ universally.

Another well-known semi-classical formulation of Hall effect is the Boltzmann theory 
within relaxation time approximation\cite{beaulac1981,allen1987,ong1991}.
If the relaxation time $\tau$ is approximated by a constant,
Hall resistivity $\rho^{\text{B}}_{xy}$ for weak magnetic field limit
does not depend on $\tau$ and is obtained as\cite{allen1987}
\begin{equation}
 \rho^{\text{B}}_{xy} = -\frac{B}{ec} \frac{\eta_{xyz}}{\eta_{xx} \eta_{yy}}
\end{equation}
where

$\eta_{ii} = \langle v_i^2 \rangle$,
$\eta_{xyz} = 
      -\left\langle v_x \left\{ v_x \frac{\partial}{\partial k_y} - v_y \frac{\partial}{\partial k_x}
        \right\} v_y \right\rangle$,
$\bm{v} = \nabla_k \varepsilon_k$ is a Fermi velocity
and $\langle \cdots \rangle $ implies an average over the Fermi surface.
For free electron dispersion $\varepsilon_{k} = \frac{k^2}{2m}$, the Boltzmann theory gives
$\rho^\text{B}_{xy} = \frac{1}{nec} B$, which coincides with
clean-limit $\sigma^\text{sc}_{xy} = -nec/B$ as $\rho^\text{B}_{xy} = -1/\sigma^\text{sc}_{xy}$. 
Therefore, it would be interesting to verify whether such correspondence holds for general band structures.
In Fig.~\ref{fig:sc}, we compare $\rho^\text{B}_{xy}$ with Chern numbers.
We found that Boltzmann theory generally agrees
with Chern numbers when the Fermi surface is essentially generated by 
a single energy band and the chemical potential is not close to van Hove singularities.
When multiple bands contribute to Fermi surface,
interband effects become important and two limits of semi-classical theory 
do not coincide.
In such case, the $\sigma_{xy}$ would sensitively depend on the details 
of scatterings or relaxations.

\begin{figure*}
  \begin{center}

    \includegraphics[width=18cm]{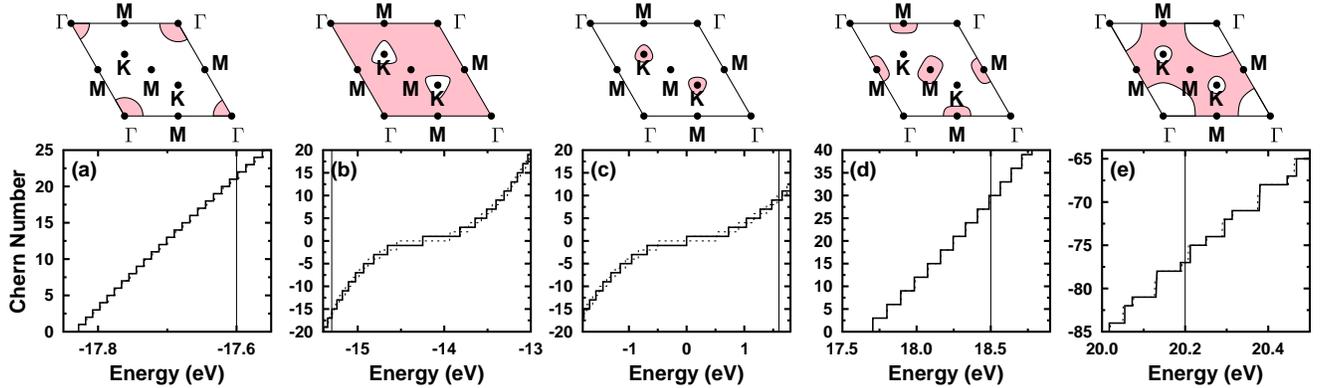}
  \end{center}
 \caption{
  (Color online)
   Chern numbers for $\phi=1/200$ at several energy regions.
   The upper panels
   show the corresponding Fermi surface calculated at the
   energies indicated by vertical lines. The shaded regions are filled by
   electrons. The dashed lines indicate the semi-classical 
   estimation $c^{\text{scq}}_F$ defined by Eq.~\ref{eq:scq}. For (b) and (c),
   the estimations by Eq.~\ref{eq:dirac} are also plotted but 
   invisible because they trace almost same points with solid lines.}
  \label{fig:sxy2}
\end{figure*}

Next, we examine the quantized plateaux of $\sigma_{xy}$ in detail.
Fig.~\ref{fig:sxy2} shows
enlarged figure of $c_F$ for $\phi=1/200$ at several energy regions. At the bottom of
energy bands $\varepsilon \sim -17.8\text{eV}$, where Fermi surface consists of an 
electron pocket around $\Gamma$ point, $c_F$
increases with a step of 1. Therefore, the quantum Hall effect in this
region resembles that of free two dimensional electron gas.

Massless Dirac fermion
regimes at $-15.7\text{eV} \lesssim\varepsilon \lesssim -13\text{eV}$ and 
$-3\text{eV} \lesssim \varepsilon \lesssim 3\text{eV}$ are shown in 
Fig.~\ref{fig:sxy2} (b) and (c).
In these regions, Fermi surface is approximated by two
massless Dirac cones centered at $K$ points. Here, two Landau bands are
grouped together and the energy gap between them are extremely
small. Well-developed energy gaps appear between these grouped Landau
bands and the Chern number $c_F$ increases with 2 between adjacent grouped
bands. This behaviour may be explained from the existence of two
equivalent Fermi surface segments. However, naive interpretation would
give $c_F = 2n$ which differs with actual steps $c_F =
2n+1$. 
These plateaux have been theoretically obtained by solving the Dirac equation under magnetic
fields\cite{gusynin2005}
or using the simplified tight-binding model for carbon $\pi$-orbital\cite{sheng2006,hatsugai2006}.
In the present calculation, we included realistic band structure which deviates 
from the linear dispersion $\varepsilon_k \sim \pm v|k|$.
The Chern numbers do not change by such details because topological number is robust as far as 
energy gaps do not close. However, the positions of steps are influenced by the band structure.
It is most clearly seen in Fig.~\ref{fig:sxy2} (c) as a violation of the electron-hole symmetry 
around $\varepsilon = 0$.
Namely, the widths of plateaux for $\varepsilon < 0$ are narrower than those for $\varepsilon > 0$.
The quantitative explanation of such behaviour will be given below.

Near $\varepsilon \sim 18 \text{eV}$, 
Fermi surface consists of three electron pockets around M points. 
We found that three Landau bands are grouped together and Chern number
can be written as $c_F = 3n$. 
More interesting behavior is realized in
other energy regions. For example, Fermi surface around $\varepsilon
\sim 20\text{eV}$ is composed of a hole pocket around $\Gamma$ point
and two hole pockets around K point. In this case, $c_F$ shows two
differnet steps with increments of 1 and 2 as presented in Fig.~\ref{fig:sxy2} (e).
This behavior can be understood 
by an assumption that $c_F$ increases 1 (2) when the semi-classical quantization 
condition to Fermi surface around $\Gamma$ (K) points is satisfied.

Let us perform quantitative comparison with the semi-classical quantization.
Onsager's semi-classical quantization condition\cite{onsager1952} can be written as

\begin{equation}
  \frac{S_i(\varepsilon)}{\Omega_\text{BZ}} = (n + \gamma) \phi.
\end{equation}

We assume that Fermi surface segments individually quantize
by above condition. When the chemical potential is located between quantized levels
\begin{equation}
  n -1 + \gamma < \frac{S_i(\mu)}{\Omega_\text{BZ}} < n + \gamma,
\end{equation}
we further assume that Hall conductance $\sigma_{xy}$ is given by a quantized value
$-\frac{e^2}{h} n$. Summing up $\sigma_{xy}$ from all Fermi surface segments, we arrive an expression
for Hall conductance
\begin{equation}
 \sigma^\text{scq}_{xy} (\mu) =   -\frac{e^2}{h} c_\text{scq}(\mu); \;
  c_\text{scq} (\mu) = \sum_i \left\lfloor \frac{ S_i(\mu)}{\phi\Omega_{\text{BZ}}}  + 1 - \gamma \right\rfloor 
  \label{eq:scq}
\end{equation}

where $\lfloor \cdots \rfloor$ denotes the largest integer not exceeding its argument.
If we choose $\gamma = \frac{1}{2}$, $c_\text{scq}$ explains  quantized Hall
conductance in several energy regions. In Fig.~\ref{fig:sxy2}, 
the $c_\text{scq}$ is plotted as dashed lines. They completely agree with $c_F$ 
for Fig.~\ref{fig:sxy2} (a) and (d). When multiple types of Fermi surface segments 
exist,  the semi-classical quantization deviates
from calculated $c_F$. It may be caused by multi-band effects.

At massless Dirac fermion regions, the $c_\text{scq}$ fails to explain $c_F$
because semi-classical quantization near massless Dirac cones is different\cite{mcclure1956}.
The existence of topological phases influences the quantization\cite{roth1966,mikitik1999,sundaram1999}.
Recently, P. Gosselin \textit{et al}\cite{pierre08} have shown
that semiclassical quantization condition of massive Dirac fermion is described by $\gamma = 0$ and $1$.
Extrapolating this result to massless limit, we obtain a modified expression of $c_\text{scq}$ 
for Dirac fermion:

\begin{equation} 
  c^\text{D}_\text{scq} (\mu) = \left\lfloor \frac{ S_D(\mu)}{\phi\Omega_{\text{BZ}}} + 1\right\rfloor + 
    \left\lfloor \frac{ S_D(\mu)}{\phi\Omega_{\text{BZ}}} \right\rfloor
    = 2 \left\lfloor \frac{ S_D(\mu)}{\phi\Omega_{\text{BZ}}} \right\rfloor + 1
  \label{eq:dirac}
\end{equation}

where $ S_D(\mu)$ is the area enclosed by a Fermi surface segment around K point.
This equation is also plotted in Fig.~\ref{fig:sxy2} (b) and (c) but almost invisible 
because it completely traces the  Chern number $c_F$. 
Thus, we found an expression
of the Hall conductance for Dirac fermion regime. 
It completely predicts the positions of quantized plateaux and explains asymmetry around $\varepsilon=0$.
If bias voltage toward graphene can be sensitively controlled, the prediction by Eq.~\ref{eq:dirac} may be 
compared with experimentally observed plateau structure.

In summary, we demonstrated that quantized Hall conductance of general band
structure can be practically calculated from topological numbers defined to
energy bands under magnetic fields.  Using graphene as an example of
multi-band models, it was shown that envelope of quantized Hall conductance
coincides with semi-classical expressions when magnetic field is sufficiently
weak. The semi-classical quantization can predict the positions of plateaux for
simple bands.  For Dirac fermion regions, modified expression was found to
predict the plateaux completely.  This expression may be  utilized for analysis
of experimental data.

This research by MA is partially supported by a Grant-in-Aid for Scientific Research, 
No.~18540331 from JSPS.
The work by YH is also supported in part by Grants-in-Aid for Scientific Research,
No. 20340098 and No. 20654034 from JSPS,
No. 220029004 (physics of new quantum phases in super clean materials)
and 20046002 (Novel States of Matter Induced by Frustration) on Priority Areas from MEXT.

\bibliography{paper}

\end{document}